\documentclass[preprnt,12pt, a4paper]{elsarticle}
\usepackage{amssymb}
\usepackage[utf8]{inputenc} \usepackage{float} \usepackage{xcolor} \restylefloat{table} \journal{SoftwareX}
\usepackage{tabularx}
\usepackage{url}
\usepackage{listings}
\usepackage{xcolor}

\definecolor{codegreen}{rgb}{0,0.6,0}
\definecolor{codegray}{rgb}{0.5,0.5,0.5}
\definecolor{codepurple}{rgb}{0.58,0,0.82}
\definecolor{backcolour}{rgb}{0.95,0.95,0.92}

\lstdefinestyle{mystyle}{
	backgroundcolor=\color{backcolour},   
	commentstyle=\color{codegreen},
	keywordstyle=\color{magenta},
	numberstyle=\tiny\color{codegray},
	stringstyle=\color{codepurple},
	basicstyle=\ttfamily\footnotesize,
	breakatwhitespace=false,         
	breaklines=true,                 
	captionpos=b,                    
	keepspaces=true,                 
	numbers=left,                    
	numbersep=5pt,                  
	showspaces=false,                
	showstringspaces=false,
	showtabs=false,                  
	tabsize=2
}

\begin{document}
\begin{frontmatter}

\title{GstLAL: A software framework for gravitational wave discovery}

\author{Kipp Cannon}
\address{RESCEU, The University of Tokyo, Tokyo, 113-0033, Japan}

\author{Sarah Caudill}
\address{Nikhef, Science Park, 1098 XG Amsterdam, Netherlands}

\author{Chiwai Chan}
\address{RESCEU, The University of Tokyo, Tokyo, 113-0033, Japan}

\author{Bryce Cousins}
\address{Department of Physics, The Pennsylvania State University, University Park, PA 16802, USA}
\address{Institute for Computational and Data Sciences, The Pennsylvania State University, University Park, PA 16802, USA}

\author{Jolien D. E. Creighton}
\address{Leonard E.\ Parker Center for Gravitation, Cosmology, and Astrophysics, University of Wisconsin-Milwaukee, Milwaukee, WI 53201, USA}

\author{Becca Ewing}
\address{Department of Physics, The Pennsylvania State University, University Park, PA 16802, USA}
\address{Institute for Gravitation and the Cosmos, The Pennsylvania State University, University Park, PA 16802, USA}

\author{Heather Fong}
\address{RESCEU, The University of Tokyo, Tokyo, 113-0033, Japan}
\address{Graduate School of Science, The University of Tokyo, Tokyo 113-0033, Japan}

\author{Patrick Godwin}
\address{Department of Physics, The Pennsylvania State University, University Park, PA 16802, USA}
\address{Institute for Gravitation and the Cosmos, The Pennsylvania State University, University Park, PA 16802, USA}

\author{Chad Hanna}
\address{Department of Physics, The Pennsylvania State University, University Park, PA 16802, USA}
\address{Institute for Gravitation and the Cosmos, The Pennsylvania State University, University Park, PA 16802, USA}
\address{Department of Astronomy and Astrophysics, The Pennsylvania State University, University Park, PA 16802, USA}
\address{Institute for Computational and Data Sciences, The Pennsylvania State University, University Park, PA 16802, USA}

\author{Shaun Hooper}
\address{University of Western Australia, Crawley, Western Australia 6009, Australia}

\author{Rachael Huxford}
\address{Department of Physics, The Pennsylvania State University, University Park, PA 16802, USA}
\address{Institute for Gravitation and the Cosmos, The Pennsylvania State University, University Park, PA 16802, USA}

\author{Ryan Magee}
\address{Department of Physics, The Pennsylvania State University, University Park, PA 16802, USA}
\address{Institute for Gravitation and the Cosmos, The Pennsylvania State University, University Park, PA 16802, USA}

\author{Duncan Meacher}
\address{Leonard E.\ Parker Center for Gravitation, Cosmology, and Astrophysics, University of Wisconsin-Milwaukee, Milwaukee, WI 53201, USA}

\author{Cody Messick}
\address{Department of Physics, The Pennsylvania State University, University Park, PA 16802, USA}
\address{Institute for Gravitation and the Cosmos, The Pennsylvania State University, University Park, PA 16802, USA}

\author{Soichiro Morisaki}
\address{Institute for Cosmic Ray Research, The University of Tokyo, 5-1-5 Kashiwanoha, Kashiwa, Chiba 277-8582, Japan}

\author{Debnandini Mukherjee}
\address{Department of Physics, The Pennsylvania State University, University Park, PA 16802, USA}
\address{Institute for Gravitation and the Cosmos, The Pennsylvania State University, University Park, PA 16802, USA}

\author{Hiroaki Ohta}
\address{RESCEU, The University of Tokyo, Tokyo, 113-0033, Japan}

\author{Alexander Pace}
\address{Department of Physics, The Pennsylvania State University, University Park, PA 16802, USA}
\address{Institute for Gravitation and the Cosmos, The Pennsylvania State University, University Park, PA 16802, USA}

\author{Stephen Privitera}
\address{Albert-Einstein-Institut, Max-Planck-Institut für Gravitationsphysik, D-14476 Potsdam-Golm, Germany}

\author{Iris de Ruiter}
\address{Astronomical Institute, Anton Pannekoek, University of Amsterdam, Science Park 904, 1098 XH Amsterdam, Netherlands}
\address{Nikhef, Science Park, 1098 XG Amsterdam, Netherlands}

\author{Surabhi Sachdev}
\address{Department of Physics, The Pennsylvania State University, University Park, PA 16802, USA}
\address{Institute for Gravitation and the Cosmos, The Pennsylvania State University, University Park, PA 16802, USA}
\address{LIGO Laboratory, California Institute of Technology, MS 100-36, Pasadena, California 91125, USA}

\author{Leo Singer}
\address{Astroparticle Physics Laboratory, NASA Goddard Space Flight Center, Mail Code 661, Greenbelt, MD 20771, USA}

\author{Divya Singh}
\address{Department of Physics, The Pennsylvania State University, University Park, PA 16802, USA}
\address{Institute for Gravitation and the Cosmos, The Pennsylvania State University, University Park, PA 16802, USA}

\author{Ron Tapia}
\address{Department of Physics, The Pennsylvania State University, University Park, PA 16802, USA}
\address{Institute for Computational and Data Sciences, The Pennsylvania State University, University Park, PA 16802, USA}

\author{Leo Tsukada}
\address{RESCEU, The University of Tokyo, Tokyo, 113-0033, Japan}
\address{Graduate School of Science, The University of Tokyo, Tokyo 113-0033, Japan}

\author{Daichi Tsuna}
\address{RESCEU, The University of Tokyo, Tokyo, 113-0033, Japan}
\address{Graduate School of Science, The University of Tokyo, Tokyo 113-0033, Japan}

\author{Takuya Tsutsui}
\address{RESCEU, The University of Tokyo, Tokyo, 113-0033, Japan}

\author{Koh Ueno}
\address{RESCEU, The University of Tokyo, Tokyo, 113-0033, Japan}

\author{Aaron Viets}
\address{Leonard E.\ Parker Center for Gravitation, Cosmology, and Astrophysics, University of Wisconsin-Milwaukee, Milwaukee, WI 53201, USA}

\author{Leslie Wade}
\address{Department of Physics, Hayes Hall, Kenyon College, Gambier, Ohio 43022, USA}

\author{Madeline Wade}
\address{Department of Physics, Hayes Hall, Kenyon College, Gambier, Ohio 43022, USA}

\begin{abstract}

The GstLAL library, derived from Gstreamer and the LIGO Algorithm Library,
supports a stream-based approach to gravitational-wave data processing.
Although GstLAL was primarily designed to search for gravitational-wave
signatures of merging black holes and neutron stars, it has also contributed to
other gravitational-wave searches, data calibration, and
detector-characterization efforts. GstLAL has played an integral role in all of
the LIGO-Virgo collaboration detections, and its low-latency configuration has
enabled rapid electromagnetic follow-up for dozens of compact binary
candidates.

\end{abstract}


\end{frontmatter}


\section{Motivation and significance}
\label{sec:intro}

Gravitational waves were originally predicted by Einstein in
1916~\cite{Einstein:1916cc} as a consequence of general relativity, which
describes gravity as the warping of space and time caused by mass and
energy~\cite{Einstein:1915ca}.  Two extremely massive objects orbiting one
another e.g., black holes or neutron stars, warp space dynamically and send
ripples across the universe that can be observed here on Earth. As they pass
by, gravitational waves stretch and squeeze the space around Earth by less than the
width of an atom compared to the Earth's diameter.  Due to their tiny effect on
scientific instruments, gravitational waves were not observed until 100 years
after their initial prediction. Technological advances in laser interferometry
led to the discovery of gravitational waves from a merging binary black hole in
2015~\cite{Abbott:2016blz}. This watershed moment was made possible by the
Advanced Laser Interferometer Gravitational-wave
Observatory (LIGO) ~\cite{TheLIGOScientific:2014jea} and the scientists of the LIGO and
Virgo Collaborations.

\begin{figure}[h!]
\includegraphics[width=\textwidth]{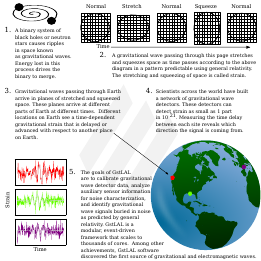}
\caption{\label{fig:info} Gravitational wave infographic.  Gravitational wave
data is time series, audio frequency data that is noise dominated.  GstLAL
identifies signals consistent with the predictions of general relativity as
measured by multiple gravitational wave detectors and assesses the probability
that these signals come from merging neutron stars and/or black holes in near
real-time.}
\end{figure}

Advanced LIGO and Advanced Virgo~\cite{TheLIGOScientific:2014jea,
TheVirgo:2014hva} are the currently operating worldwide network of
kilometer-scale laser interferometric gravitational wave observatories which
have measured gravitational wave signals. These detectors provide a new way to
observe our Universe and enable a vast amount of new science. LIGO's
observations have already deepened our understanding of the populations of
compact objects such as neutron stars and black
holes~\cite{LIGOScientific:2018mvr, LIGOScientific:2018jsj}, and they have also
offered new tests of fundamental physics~\cite{Abbott:2017xzu, Abbott:2018lct,
LIGOScientific:2019fpa, Abbott:2019yzh}.  The strong gravity regime probed by
compact binary mergers is a laboratory for novel tests of general relativity,
and a joint observation of gravitational waves along with electromagnetic
waves~\cite{GBM:2017lvd} has taught us how matter behaves in the most extreme
conditions~\cite{Abbott:2018exr, Nicholl:2017ahq}. 

The science made possible by LIGO and Virgo is reliant on measuring
miniscule changes in the arm lengths of the interferometers known
as strain. The perturbations caused by incident gravitational waves manifest
themselves as variations in the intensity of laser light output.
\emph{Detector calibration} aims to accurately map the intensity of the output
to differential changes in the arm length through real-time signal-processing.
The calibrated strain data contains the encoded properties of the astrophysical
systems that produce gravitational waves.  The analysis of this data is
complicated by the presence of a vast array of transient noise sources.
\emph{Detector characterization} aims to quantify departures from stationary
noise to identify times where instrumental issues are so severe that the data
should not be analyzed or there may be a coupling between environmental sensors
(such as seismometers) and the gravitational wave strain data. Once data is
calibrated and assessed for quality, it is analyzed by a host of detection
algorithms to identify potential gravitational wave signals. In many cases the
signals are invisible to the naked-eye in raw data and discovering them
requires sophisticated techniques that may involve checking millions of models
against each segment of data.  All three of these activities require
substantial cyberinfrastructure.  The GstLAL software framework~\cite{gstlal}
was initially designed to support low-latency compact binary searches to
facilitate multi-messenger astronomy, but since its conception it has grown to
be a key component of the software used to produce accurately calibrated strain
data~\cite{Viets:2017yvy}, and recently it has contributed to detector
characterization
efforts~\cite{idq-methods, idq-in-gstlal}. The GstLAL framework is now
contributing key cyberinfrastructure to all three of these key aspects of
gravitational wave data analysis.  This paper will describe how the GstLAL
software is used in gravitational-wave
searches~\cite{Messick:2016aqy,Sachdev:2019vvd}, provide examples, and describe
the history and impact of GstLAL on gravitational wave discovery.

\section{Software description}
\label{sec:description}

Gravitational wave strain data quantifies how the distance between two points
will change as a gravitational wave passes. The current gravitational wave
observatories are sensitive to changes in strain and measure the stretching and
squeezing of space as a function of time.  The Advanced
LIGO~\cite{TheLIGOScientific:2014jea} and Advanced
Virgo~\cite{TheVirgo:2014hva} gravitational wave detectors are most sensitive
to strain frequencies between 10Hz--10kHz, which is remarkably close to the
frequency range of the human ear~\cite{olson1967music}. For this reason, there
is a close connection between the analysis of gravitational wave data and the
analysis of audio data.  Indeed, techniques such as low pass filtering, high
pass filtering, channel mixing, and gating apply equally well to both audio
processing and gravitational wave data processing, which provides the
motivation for basing GstLAL on Gstreamer~\cite{gstreamer}.

Gstreamer~\cite{gstreamer} is an open-source, cross-platform multimedia
processing framework designed to execute audio and video processing graphs
organized into three basic elements: sources, filters and sinks, which are
provided by dynamically loaded plugins.  A valid Gstreamer graph,
called a pipeline, connects elements together ensuring that the capabilities of
each element are satisfied.  The data are passed along in \emph{buffers} that
store both the memory location of the raw data as well as rich metadata.
Pipelines can be used to construct complex workflows and scale to thousands of
elements. GstLAL combines standard Gstreamer signal processing elements with
custom elements to analyze LIGO strain data.
\begin{figure}
\includegraphics[width=\textwidth]{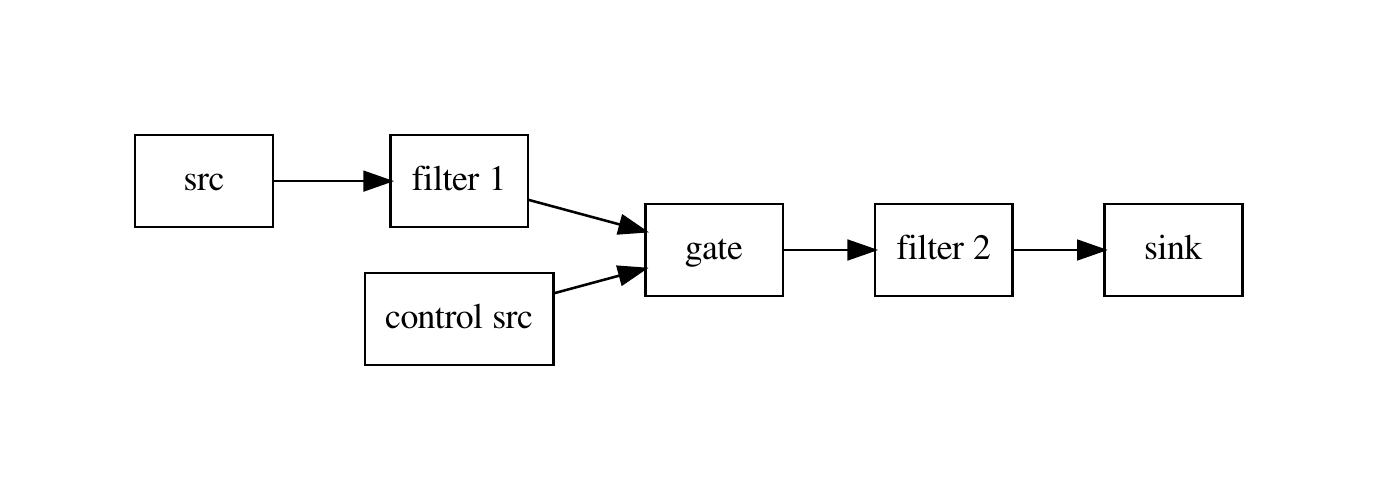}
\caption{\label{fig:gst}A basic Gstreamer graph.  Data starts at a source,
``src", e.g., a file on disk or a network socket, and then is passed through a
filter element, ``filter 1", which transforms the data, e.g., by performing a
low pass filter.  A second data stream starts from ``control src" and the
output of filter 1 is moderated by a gate controlled by the state of ``control
src".  The output of the gate is filtered through ``filter 2" and sent to a
sink which could be another file on disk or a network socket.}
\end{figure}

The GstLAL software began development in 2008 through the exploration of novel
techniques for filtering gravitational wave data~\cite{Cannon:2010qh}.  It
derives its name from ``Gstreamer wrappings of the LIGO Algorithm Library\footnote{\url{https://git.ligo.org/lscsoft/lalsuite}}
(LAL)". GstLAL began to take on its modern form by 2009 and has been since
actively developed as open source software. GstLAL currently resides in the
LIGO Scientific Collaboration hosted GitLab instance at
\url{https://git.ligo.org/lscsoft/gstlal}~\cite{gstlal}.  

GstLAL is primarily a mix of Python and C with contributions from $75$ authors
distributed across North America, Europe, Asia and Australia.  The master
branch currently has over 13,000 commits and 250,000 lines of code.  GstLAL is
released under the GPLv2 license with 44 distinct releases since
2011~\cite{ligo-software}.  In 2012, code solely used for gravitational wave
searches for compact binaries was split off into its own package:
GstLAL-Inspiral. GstLAL-Inspiral has had 45 distinct releases since then.  In
2014, code used for gravitational wave burst detection was split into its own
package, GstLAL-Burst, with nine distinct releases. And finally, in 2014 code
used primarily for LIGO strain data calibration was split into its own package,
GstLAL-Calibration, with 60 releases. In addition to tar-ball releases, RedHat
and Debian compatible packages were produced for the LIGO Data Grid reference
platforms~\cite{ligo-packages}.

At present, Docker containers with the full GstLAL/LALSuite software stack are
built and distributed using the LIGO Container Registry~\cite{ligo-containers}.
The containers are built on top of the CentOS-based Scientific Linux 7, which
currently serves as the reference operating system on the LIGO Data Grid.
Binary executables are linked against Intel's high-performance Math Kernel
Library (MKL), and compiled to leverage Advanced Vector Extensions. Optimized
versions of the GstLAL software stack tuned at the compiler-level to best
leverage the native features of the local computing environment are often
custom-built by users. Previous studies have demonstrated a $\gtrsim 2$ times
increase in overall code throughput as a result of software tuning.  GstLAL is
also available through CondaForge~\cite{conda-forge}. 

\subsection{Software Architecture}
\label{ssec:architecture}

Prior to 2015, gravitational waves had not been directly
observed~\cite{Abbott:2016blz}.  Although analysis techniques had been studied
for decades~\cite{Finn:1992xs, Allen:2005fk}, the character of gravitational
wave data evolved with the interferometers detectors during the initial
operation of LIGO and Virgo from 2002--2010~\cite{Abbott:2003pj, Abbott:2005pe,
Abbott:2007xi, Abbott:2009tt, Abbott:2009qj, Colaboration:2011np}.  Therefore,
we aimed to make the GstLAL software modular and easy to adapt to the
challenges of Advanced LIGO and Virgo data.  

The joint observation of gravitational waves and electromagnetic signals, known
as multi-messenger astronomy, was a significant goal for Advanced LIGO and
Advanced Virgo.  In this scenario, gravitational-wave observations were
expected to be followed up by observations with telescopes across the
electromagnetic spectrum hoping to catch a short-lived transient light source.
Discovering gravitational waves quickly is critical because electromagnetic
counterparts may quickly fade.  GstLAL was designed to offer analysts an
extremely short time-to-solution to help ensure that electromagnetic
counterparts could be observed quickly.  

The key design principles of GstLAL are:
\begin{description}
\item{\textbf{Plugin-based:}} Libraries within GstLAL provide Gstreamer
plugins to perform gravitational wave specific signal processing tasks. These
are mixed together with stock Gstreamer plugins to produce gravitational wave
analysis pipelines.  Plugins provide elements, which are the building blocks of
signal processing workflows.  These elements can be ordered in multiple ways
with minimal coding effort which allows for quick exploratory work and
development of new methods.
\item{\textbf{Streaming:}}  The goal of Gstreamer is to provide ultra low
latency signal processing suitable for audio and video playback and editing.
GstLAL pipelines typically work with streaming data buffers $\lesssim 1$ s in
duration.
\item{\textbf{Event-driven:}} GstLAL analysis pipelines are designed to run
continuously as data is collected.  Each application runs an event loop which
controls both application level operations as well as settings within a given
plugin.  This allows for the control of program behavior to be altered while
the application is running.  Dynamic program control is facilitated through
embedding the microservices framework Bottle~\cite{bottle}.  Simple http APIs
push information to the program, retrieve information or alter the program's
behavior.
\item{\textbf{Scalable:}} The GstLAL framework is designed both to scale to
dozens of cores on a single computer using the multi-threading provided by
Gstreamer and to scale to thousands of cores across a compute cluster
leveraging HTCondor directed acyclic graph (DAG) scheduling~\cite{condor}.
\end{description} 


The GstLAL project is currently comprised of five distinct packages. 1)
\texttt{gstlal}, 2) \texttt{gstlal-ugly}, 3) \texttt{gstlal-inspiral}, 4)
\texttt{gstlal-calibration} and 5) \texttt{gstlal-burst}, all of which are
described below.

\subsection{{\upshape \texttt{gstlal}} package}
\label{ssec:gstlal}

The \texttt{gstlal} package curates a set of core plugins, functions and
applications that are used by nearly every analysis workflow developed within
GstLAL. The \texttt{gstlal} package is a dependency for all of the remaining
packages which are described in subsequent sections.  The \texttt{gstlal}
package provides Gstreamer elements for Finite-Impulse-Response filtering
(\texttt{lal\_firbank}), $N\to M$-channel matrix operations
(\texttt{lal\_matrixmixer}), data whitening (\texttt{lal\_whiten}), and data
gating (\texttt{lal\_gate}).  The \texttt{gstlal} package also provides basic
python APIs for building Gstreamer pipelines in the module \texttt{pipeparts},
basic data access routines in the module \texttt{datasource}, and a base class
for event handling in the module \texttt{simplehandler}.  

\subsection{{\upshape \texttt{gstlal-ugly}} package}
\label{ssec:gstlal-ugly}

The \texttt{gstlal-ugly} package is an incubator package for software that is
in development. Eventually all \texttt{gstlal-ugly} software is migrated to one
of the other packages.

\subsection{{\upshape \texttt{gstlal-inspiral}} package}
\label{ssec:gstlal_inspiral}

The primary purpose of the \texttt{gstlal-inspiral} package is to house the
GstLAL-based search for compact binaries~\cite{Messick:2016aqy,
Sachdev:2019vvd}, which centers around the application
\texttt{gstlal\_inspiral}.  The GstLAL-based compact binary pipeline was
created to make near real-time gravitational-wave detections and aimed to one
day detect electromagnetically bright systems before
coalescence~\cite{Cannon:2011vi,Sachdev:2020lfd}.

The GstLAL-based compact binary search is a matched-filter search that
incorporates efficient time-domain filtering~\cite{Cannon:2010qh,
Cannon:2011tb, Cannon:2011xk, Cannon:2011rj, Cannon:2012gq, Smith:2012du,
Cannon:2011vi, Tsukada:2017cuf} of a set of template waveforms that match the
gravitational wave signals of merging black holes and neutron
stars~\cite{Mukherjee:2018yra}.  LIGO and Virgo detectors are prone to bursts
of nonstationary noise called glitches~\cite{TheLIGOScientific:2016zmo} and
determining the difference between gravitational waves and glitches is well
suited for many classification algorithms. GstLAL Inspiral implements a
classification scheme that is a hybrid of hypothesis testing techniques with
some elements coming from machine learning approaches.  The classifier is an
approximate likelihood ratio comprised of many terms which began as a custom
implementation of Naive Bayes classification~\cite{zhang2004optimality} applied
to gravitational wave searches~\cite{Cannon:2008zz}.  It was realized early in
the project that two things were apparent.  First, it wasn't practical to treat
the classifier as entirely data driven relying purely on training sets.
Training sets to adequately classify the full parameter space were too
expensive to produce.  Second, correlations between some parameters had to be
tracked in order to classify well.  The first point was addressed by developing
semi-analytic models to describe parts of the data~\cite{Hanna:2019ezx} and the
second point was addressed by factoring the multi-dimensional likelihood ratios
into groups of lower dimensional, but not one dimensional,
distributions~\cite{Cannon:2015gha, Cannon:2012zt}.

The GstLAL-based compact binary search has two modes.  The first is a
near-real-time, ``low-latency" mode that discovers and reports compact binaries
within tens of seconds of the signal arriving at Earth.  The second is an
``offline" mode that efficiently processes data in batch jobs where
time-to-solution is not as important as computational efficiency and
reproducibility.  Although both modes share $\gtrsim 95\%$ of the same code,
their behavior and design are very different in order to address the differing
concerns of real-time vs. batch processing.

The low-latency mode is a collection of typically $\mathcal{O}(1000)$
microservices that communicate (modestly) with one another asynchronously
through http using python Bottle, through an Apache Kafka queue, and through a
shared file system.  Each one of these microservices processes a portion of the
nearly 2 million models used in the current low-latency compact binary search.
The low-latency workflow is designed to be fault tolerant. If a job dies,
another is restarted to take its place.  Since information is exchanged
asynchronously and there is no guarantee of job success, the behavior in this
mode is non-deterministic.  In contrast, the offline mode has a fully
deterministic execution which can be reproduced to floating point precision.
The determinism is imposed by organizing each job in a directed acyclic graph
(DAG) using HTCondor.

\subsection{{\upshape \texttt{gstlal-burst}} package}
\label{ssec:gstlal_burst}

The GstLAL-based burst package is a collection of utilities intended to
search for gravitational-wave sources other than compact binaries as well
as non-astrophysical noise transients. One of the
recent developments is a pipeline searching for cosmic strings, which are
hypothetical objects considered to have formed in the early universe. The
pipeline uses time-domain stream-based signal processing algorithms, along
with a classification scheme using parameters specific to the search.
The algorithms used are mostly in common with the \texttt{gstlal-inspiral}
package, but simplified due to the smaller number of templates required for
matched filtering.

In addition, \texttt{gstlal-burst} provides utilities to identify and extract
features from non-Gaussian noise transients in near real-time ($O(5 s)$) via the
Stream-based Noise Acquisition and eXtraction, or SNAX toolkit. The SNAX toolkit
also leverages time-domain signal processing, but instead utilizes a sine-Gaussian
basis to identify the presence of and extract features from many types of
non-Gaussian noise in strain and auxiliary data. Its main data product is
multivariate time-series data containing the extracted features, including
SNR and phase information as well as the waveform parameters of interest.

\subsection{{\upshape \texttt{gstlal-calibration}} package}
\label{ssec:gstlal_calibration}

The \texttt{gstlal-calibration} package houses the unique software used for calibration of the LIGO strain data.  Software in the \texttt{gstlal-calibration} package produces the only official LIGO strain data product used in all subsequent analysis.  Calibration of LIGO strain data involves standard signal processing and digital filtering techniques in order to derive the differential arm motion observed in the LIGO detectors from the detector's digital readouts \cite{Abbott:2016jsd, Viets:2017yvy, Tuyenbayev2016}.  Many of the signal processing and digital filtering plugins used by the calibration pipeline \texttt{gstlal\_compute\_strain} are housed in the \texttt{gstlal} or \texttt{gstlal-ugly} packages.  A few plugins unique to the calibration process as well as calibration-specific python APIs are housed in the \texttt{gstlal-calibration} package.

Much like the GstLAL-based compact binary pipeline, the LIGO calibration pipeline is built to operate in two modes: a ``low-latency" mode and an ``offline" mode.  The low-latency LIGO calibration pipeline operates on hardware located physically at the two LIGO detector sites, LIGO Hanford  in Hanford, WA and LIGO Livingston in Livingston, LA.  This pipeline produces calibrated LIGO data and a bit-wise state-vector that indicates the fidelity of the calibrated data within $O(5 s)$ for each detector at the respective detector sites.  The low-latency calibration process involves using a combination of digital filtering performed in the LIGO front-end computers, which are directly connected to the LIGO detectors and employ the CDS Real-time Code Generator (RCG) core software \cite{RCGReleaseNotes}, and further digital filtering and processing performed by the GstLAL calibration software running on non-front-end hardware located at the LIGO detector sites.  This two-step process takes advantage of the access the front-end computing system has to the installed detector filters and models and the advanced stream-based filtering techniques housed in the GstLAL software packages \cite{Viets:2017yvy}.

There is often a need to re-calibrate the strain data after the initial low-latency data calibration in order to improve calibration accuracy based on more sophisticated modeling or to remove systematic errors present in the low-latency calibrated strain data \cite{Viets:2017yvy, Cahillane2017, Sun2019}.  The re-calibration of LIGO data is performed using the offline mode of the \texttt{gstlal\_compute\_strain} pipeline.  In this mode, the entire calibration process is performed by software housed in the GstLAL software packages.  The offline calibrated data is processed in batch jobs using HTCondor in order to optimize computational efficiency and is completely reproducible to floating-point precision.  All analyses derived from LIGO strain data use the calibrated data produced either by the low-latency GstLAL calibration pipeline or the offline GstLAL calibration pipeline.

\section{Illustrative Examples}
\label{sec:examples}

\subsection{Example Gstreamer pipeline with GstLAL}

The following example was run on a newly instantiated CentOS 7 64 bit virtual
machine with miniconda~\cite{miniconda} installed by doing:
\begin{lstlisting}[language=Bash]

wget https://repo.anaconda.com/miniconda/Miniconda3-latest-Linux-x86_64.sh
bash Miniconda3-latest-Linux-x86_64.sh
conda create -n myenv python=2.7
conda activate myenv
conda install -c conda-forge gstlal-inspiral=1.7.3

\end{lstlisting}
To verify that it works, try:
\begin{lstlisting}[language=Bash]

$ gst-inspect-1.0 gstlalinspiral | head -n4
Plugin Details:
  Name                     gstlalinspiral
  Description              Various bits of the LIGO Algorithm Library wrapped in gstreamer elements
  Filename                 <your miniconda path>/lib/gstreamer-1.0/libgstgstlalinspiral.so

\end{lstlisting}
Next we will try a very simple Gstreamer pipeline that uses two GstLAL
elements: \texttt{lal\_peak} and \texttt{lal\_nxydump}, along with additional
Gstreamer elements to construct a pipeline that generates 10 Hz Gaussian, white
noise, finds the peak sample every second and streams the result to the
terminal screen as ASCII text.  It is possible to construct simple pipelines
such as this without any code using the Gstreamer tool
\texttt{gst-launch}~\cite{gst-launch}:
\begin{lstlisting}[language=Bash]

$ gst-launch-1.0 audiotestsrc wave=9 ! capsfilter caps=audio/x-raw,rate=10 ! lal_peak n=10 ! lal_nxydump ! filesink location=/dev/stdout

\end{lstlisting}
The first element, \texttt{audiotestsrc} is a Gstreamer element that can
provide many test signals.  The \texttt{wave=9} property sets it to be unit
variance white noise. The second element, \texttt{capsfilter} specifies that we
want the format of the output to be floating point audio data with a sample
rate of 10 Hz. Next, \texttt{lal\_peak} is the first GstLAL element. In this
example it is configured to find the largest absolute value of the signal every
10 sample points.  \texttt{lal\_nxydump} is the second GstLAL element which
converts the time-series data to two column ASCII text.  Finally,
\texttt{filesink} dumps the ASCII output to standard out.  You should see the
following output (with variations caused by the fact that the data is random):
\begin{lstlisting}[language=Bash]{GstLaunchOut.sh}

$ gst-launch-1.0 audiotestsrc wave=9 ! capsfilter caps=audio/x-raw,rate=10 ! lal_peak n=10 ! lal_nxydump ! filesink location=/dev/stdout | head -n 15
Setting pipeline to PAUSED ...
Pipeline is PREROLLING ...
Pipeline is PREROLLED ...
Setting pipeline to PLAYING ...
New clock: GstSystemClock
0.000000000     0
0.100000000     0
0.200000000     0
0.300000000     0
0.400000000     0
0.500000000     0
0.600000000     0.95523328
0.700000000     0
0.800000000     0
0.900000000     0

\end{lstlisting}
you can see that the maximum sample point was chosen in the 10 sample interval
on line 13. Other values are set to 0.

\texttt{gst-launch} is useful tool for quickly testing a simple pipeline, or
debugging, however it is not suitable for writing large applications with many
elements or situations where program control is exposed dynamically to the
user.  For building applications, GstLAL relies on the Python bindings for
Gstreamer and adds a substantial amount of gravitational wave specific
application code written in python.  An example
of the pipeline above written in the style of GstLAL applications is below.
\begin{lstlisting}[language=Python]{demo.py}

# boiler plate Gstreamer imports
import gi
gi.require_version('Gst', '1.0')
from gi.repository import GObject, Gst
GObject.threads_init()
Gst.init(None)

# Gstlal imports
from gstlal import datasource
from gstlal import pipeparts
from gstlal import simplehandler

# initialize an event loop, a pipeline and an event handler
mainloop = GObject.MainLoop()
pipeline = Gst.Pipeline("softwarex_demo")
handler = simplehandler.Handler(mainloop, pipeline)

src = pipeparts.mkaudiotestsrc(pipeline, wave = 9)
src = pipeparts.mkcapsfilter(pipeline, src, caps = "audio/x-raw, rate=10")
src = pipeparts.mkpeak(pipeline, src, n = 10)
src = pipeparts.mknxydumpsink(pipeline, src, "/dev/stdout")

if pipeline.set_state(Gst.State.PLAYING) == Gst.StateChangeReturn.FAILURE:
        raise RuntimeError("pipeline failed to enter PLAYING state")
mainloop.run()

\end{lstlisting}
We have found that, using python to procedurally build Gstreamer graphs, we can
construct enormous pipelines containing tens of thousands of distinct elements.
A prime example of this is our workhorse signal processing pipeline used for
discovering compact binary mergers as described in the next section.

\subsection{Compact binary searches}
\label{ssec:cbc_searches}

\texttt{Makefile.softwarex\_test} provides an example of the general workflow
involved in offline gravitational wave analyses, which primarily rely on the
\texttt{gstlal} and \texttt{gstlal-inspiral} packages. The miniconda
installation of GstLAL will run this example in $\sim$30 minutes on a single
machine. Production level compact binary searches analyzing data from the three advanced interferometers, on the other hand, take $\sim$1 week when distributed over $\mathcal{O}$(1000) core computing clusters with optimized software builds. The test Makefile is
entirely self-contained; the target and dependency relationships and brief
comments within describe the workflow. Although the structure presented in this
test is linear in nature, full scale searches for compact binaries are heavily
parallelized to take advantage of DAG scheduling. The requisite commands to run
this test analysis are shown below.
\begin{lstlisting}[language=Bash]

mkdir workflow-test && cd workflow-test
export LAL_PATH=${CONDA_PREFIX} GSTLAL_FIR_WHITEN=0 TMPDIR=/tmp
wget https://dcc.ligo.org/public/0168/P2000195/004/Makefile.softwarex_test
make -f Makefile.softwarex_test

\end{lstlisting}

\section{Impact}
\label{sec:impact}


GstLAL has played an integral role in the history of gravitational wave
detections, and has participated in gravitational wave searches since
S5~\cite{wade2015}.  Although GstLAL was designed specifically for
near-real-time applications, the low-latency pipeline was prevented from
searching for binary black holes at the start of Advanced LIGO's first
observing run (O1). The pipeline was only allowed to use a template bank sensitive
to electromagnetically binary neutron stars and neutron star - black hole
binaries. The software was fully capable of detecting binary black holes in near-real-time. The LIGO
collaboration desired blind binary black hole (BBH) analyses, and since BBH
systems are not expected to produce electromagnetic radiation there was no
perceived need to detect them in near-real-time. The restriction on the allowed
template bank rendered GstLAL unable to detect GW150914 in low-latency, though
it was one of two matched filter pipelines used to analyze archival data and
validate the event~\cite{Abbott:2016blz}. GW150914 was initially detected in
low-latency by the weakly-modeled burst pipeline CWB~\cite{Klimenko:2015ypf},
which demonstrated that there could be no truly blind analysis while
low-latency burst pipelines produced alerts. As a result, in late 2015 the
GstLAL pipeline was approved to include BBHs with total mass $\lesssim 100
M_\odot$ in its low-latency configuration. GstLAL quickly demonstrated its
ability to recover BBHs in low-latency; it became the first matched-filter
pipeline using waveforms based on general relativity to make a near real-time
detection of a compact binary with the discovery of
GW151226~\cite{Abbott:2016nmj}.

Development work between Advanced LIGO's first and second observing runs
focused on enabling single detector discoveries and
incorporating data from Virgo in the analysis.  
These efforts were rewarded by August 2017 as GstLAL became the first three-detector
matched-filter search in the Advanced LIGO era and, more notably, the first
(and to date, the only) gravitational-wave detection pipeline to observe a
binary neutron star merger in low-latency~\cite{GCN24168,G298048}.  Although
both Advanced LIGO interferometers and the Advanced Virgo interferometer were
operating at the time of GW170817, it was initially observed in a single
interferometer. This marked the first single detector observation of a gravitational
wave, and the autonomous identification of the candidate enabled rapid offline
follow-up of the candidate within the LIGO-Virgo collaboration. 
 
Advanced LIGO's third observing run (O3) marked the beginning of open
public alerts (OPA)\cite{GCN24045}. For the first time, candidates with
false-alarm-rates below 1 per month\footnote{after applying a trials factor to
account for the number of concurrently running searches.} were made public at
the time of discovery. Although the first public alert was distributed in error
by the collaboration~\cite{GCN24109}, the identification of binary black hole
candidate S90408an~\cite{GCN24069} marked the successful start of the era of
automated public alerts. At first only candidates appearing in two or more
detectors were approved for automated release, but GW170817 had already demonstrated
the importance of single detector searches. Indeed, two weeks into the
third observing run GstLAL was the only pipeline to detect GW190425 in
near-real-time~\cite{GCN24168,Abbott:2020uma}, further highlighting the
necessity of single detector searches. 

Two months into the observing
run, GstLAL became the only pipeline approved to release single detector
candidates as OPAs. This was a high risk, high reward endeavor. Matched filter
searches have traditionally been able to suppress the background by demanding
coincidence across interferometers; single detector candidates do not benefit
from this effect and can therefore be more susceptible to short term noise
transients. Unfortunately, this resulted in several retractions throughout O3
as the GstLAL team worked on ways to mitigate the effects of noise transients
in single detectors. By the end of the second half of the observing run,
pipeline tuning had reduced the rate of retractions.
 
GstLAL has contributed to all gravitational-wave
discoveries published by the LIGO Scientific
Collaboration~\cite{Abbott:2020uma,LIGOScientific:2018mvr,TheLIGOScientific:2017qsa,Abbott:2017oio,Abbott:2017gyy,Abbott:2017vtc,Abbott:2016nmj,Abbott:2016blz},
but it has also contributed to searches for as yet undetected sources.
Sub-solar mass and intermediate mass black holes both pose problems for
conventional models of stellar evolution, and GstLAL has directly contributed
to searches of
both~\cite{Abbott:2018oah,Salemi:2019ovz,Authors:2019qbw,Abbott:2019prv}.
Although these searches have not yet yielded any detections, the null results
have been able to place strict limits on the abundance of such objects and have
also provided the tightest limit to date on a primordial black hole model of
the dark matter. 

 
\section{Conclusions}
\label{sec:conclusions}

The GstLAL library has significantly impacted the progress of
gravitational-wave astrophysics, not only via compact binary searches, but also
through contributions to detector calibration and characterization efforts. The
low-latency GstLAL based inspiral pipeline was instrumental in the first
multi-messenger discovery with gravitational waves, and strives to lead the
march towards more remarkable observations with ground-based interferometers.

\section{Conflict of interest}

No conflict of interest exists: we wish to confirm that there are no known
conflicts of interest associated with this publication and there has been no
significant financial support for this work that could have influenced its
outcome.

\section*{Acknowledgments}
\label{}

Funding for this work was provided by the National Science Foundation through
awards: PHY-1454389, OAC-1642391, PHY-1700765, OAC-1841480, PHY-1607178, and
PHY-1847350.  Funding for this project was provided by the Charles E. Kaufman
Foundation of The Pittsburgh Foundation. Computations for this research were
performed on the Pennsylvania State University’s Institute for Computational
and Data Sciences Advanced CyberInfrastructure (ICDS-ACI) and VM hosting.  We
are grateful for computational resources provided by the Leonard E Parker
Center for Gravitation, Cosmology and Astrophysics at the University of
Wisconsin-Milwaukee.  Computing support was provided by the LIGO Laboratory
through National Science Foundation grant PHY-1764464.  GstLAL relies on many
other open source software libraries; we gratefully acknowledge the development
and support of NumPy~\cite{numpy}, SciPy~\cite{scipy}, PyGTK~\cite{pygtk},
PyGST~\cite{pygst}, Bottle~\cite{bottle}, Kafka~\cite{kafka},
Fftw3F~\cite{fftw3f}, Intel MKL~\cite{intelmkl}, GLib2~\cite{glib2}, GNU
Scientific Library~\cite{gsl}, and GWpy~\cite{gwpy}.  The authors gratefully
acknowledge the LIGO-Virgo-Kagra collaboration for support, review, and
valuable critiques throughout various stages of development of the GstLAL
library. We are especially thankful for collaborations within the Compact
Binary Coalescence working group.


\bibliographystyle{elsarticle-num} 
\bibliography{references}



\section*{Current code version}
\label{}


\begin{table}[h]
\begin{tabular}{|l|p{6.5cm}|p{6.5cm}|}
\hline
\textbf{nr.} & \textbf{code metadata description} &  \\
\hline
c1 & current code version & 1.7.3 \\
\hline
c2 & permanent link to code/repository used for this code version & $https://git.ligo.org/lscsoft/gstlal$ \\
\hline
c3 & code ocean compute capsule & N/A\\
\hline
c4 & legal code license   & gnu general public license \\
\hline
c5 & code versioning system used & git \\
\hline
c6 & software code languages, tools, and services used & python, c, sqlite \\
\hline
c7 & compilation requirements, operating environments \& dependencies & scientific linux 7, python 2.7 \\
\hline
c8 & if available link to developer documentation/manual & $https://lscsoft.docs.ligo.org/gstlal/$ \\
\hline
c9 & support email for questions & gstlal-discuss@ligo.org \\
\hline
\end{tabular}
\caption{code metadata (mandatory)}
\label{} 
\end{table}

\end{document}